# Elastic wave velocities under methane hydrate growth in Bentheim sandstones

Mathias M. Sæther[1], Per Lunde[1,3], Stian Almenningen[2], Geir Ersland[2]

[1] Acoustics group, Department of Physics and Technology, University of Bergen, Postboks 7803, N-5020 Bergen, Norway
[2] Petroleum and process technology group, Department of Physics and Technology, University of Bergen, Postboks 7803, N-5020 Bergen, Norway
[3] NORCE, P.O. Box 6031 Postterminalen, N-5892 Bergen, Norway
Contact email: msa022@uib.no

**Abstract**

Experimental acoustic laboratory measurement methods for hydrate-bearing poroelastic solid media are biefly reviewed. A measurement example using the Fourier spectrum method is given, for compressional and shear wave velocities in hydrate-bearing Bentheim sandstone.

## 1  Introduction

Methane gas hydrates exist in vast quantities beneath the ocean-bed, inland seas, and in the permafrost, and may potentially be used as an energy source [1,2,3]. The high energy density and the relatively low $CO_2$-footprint (compared with coal) makes utilizing these hydrate resources highly relevant. Hydrate deposits stretch over large areas and monitoring hydrate deposits must therefore rely on remote monitoring methods. Acoustic parameters such as compressional ($c_P$) [4,5,6,7] and shear ($c_S$) wave velocities [6,7] and the compressional ($\alpha_P$) [7] and shear ($\alpha_S$) wave attenuation coefficients [7] are known to depend on the hydrate saturation ($S_H$). Acoustic methods have been pointed to as candidates for remote detection and monitoring of hydrate deposits [1].

The acoustic parameters $c_P$, $c_S$, $\alpha_P$ and $\alpha_S$ of hydrate-bearing poroelastic solid media do not only depend on $S_H$ but on many other factors, such as the free gas saturation ($S_g$), water saturation ($S_w$), measurement frequency, sediment composition and hydrate formation pattern. To understand field data such as acoustic well-log measurements [8,9], information on how all these listed factors affect the acoustic parameters is needed. Some of his information may be gained in the laboratory, where these factors may be adjusted and controlled [4,5,6,7]. Bentheim sandstone is a relatively homogeneous material and is fairly well described in the literature [23]. Thus, Bentheim sandstone was chosen in this study as the laboratory "host sediment" for hydrate to grow within.

Both sonic [6,7] (typically < 30 kHz) and ultrasonic [4,5,10,11,12] (typically >100 kHz) frequencies can be used to study elastic wave velocities in methane hydrate bearing samples in the laboratory. For sonic frequencies, resonance methods have been used as measurement methods [6,7]. In studies using ultrasonic measurement frequencies, the time of flight to the first arrival of the acoustic pulse transmitted through the hy-drate-bearing samples has been measured [4,5,10,11,12]. By using calibration rods of known dimensions and material parameters, the inherent time of flight in the transducers and in the electronics in the experimental setup can be subtracted [10]. Using this method, $c_P$ and $c_S$ have been measured during hydrate growth for hydrate-bearing samples in the laboratory.

At the first arrival of the signal, the signal strength is low and noise may affect the transit time measurements. There is no defined frequency in the transient of the signal and the method does not allow for attenuation measurements. When using shear wave transducers, both P-waves and S-waves are generated. It is apparent from for example Ref. [10] that P-wave components may interfere with the S-wave arriving later in the time trace. The accuracy of measuring the first arrival of the signal has been debated [13,14].

Another signal processing method is the Fourier spectrum method which has been used to measure elastic wave attenuation and velocity spectra [15,16]. The method relies on taking the Fourier transform of short pulses and does therefore not depend on accurate time domain measurements. In such methods where the frequency is well-defined, diffraction effects may also be corrected for [17,18].

In previous laboratory studies on hydrate-bearing poroelastic media, the hydrate formation pattern has received considerable attention [4,5,10,11,12]. The hydrate content has been estimated by interpreting the measured $c_P$ and $c_S$ using numerical models [19,20]. It has been investigated whether (i) hydrates form primarily within the pore fluid, or (ii) hydrates form and grow on individual grains, becoming a part of the frame, or (iii) hydrates form and grow at and around grain contacts, becoming part of the frame but also cementing the grains together. A general trend is that methane hydrates tend to form in the pore fluid in systems with high water content and adhere to the solid frame in systems where the water content is low [4,5,10,11,12]. Tohidi et al. [21] showed that methane hydrate forms primarily in the center of pores in "water-rich" systems initially containing bubbles of gas. Waite et al. [22] found that hydrate cements sediment grains in "gas-rich" systems initially containing discrete units of water. Apart from studies where synthetic hydrates (THF) have been formed, there appear to be few laboratory studies of the elastic properties of hydrate bearing consolidated poroelastic media; only one such laboratory study has been identified by the present authors [10].

In both consolidated poroelastic media [10] and unconsolidated [6,7,4,5,11,12] sand, a clear increase of both $c_P$ and $c_S$ as a function of $S_H$ has been found. This increase is found to be higher for unconsolidated sand. This indicates that different elastic properties should be expected for the same amount of methane hydrate depending on whether the reservoir sediment is a consolidated porous rock or unconsolidated sandy sediment. An example of such a consolidated porous rock is Bentheim sandstone, which has been used in this study.

## 2 Elastic wave velocities for hydrate-bearing Bentheim sandstone using the Fourier spectrum signal processing method

In an experimental study described by Sæther [17], acoustic properties of hydrate-bearing Bentheim sandstone was investigated e.g. using the Fourier spectrum signal processing method. Shear wave transducers were designed and constructed. The transducers were excited with short pulses and both P-waves and S-waves were measured. The measurement method, together with numerical simulation models for acoustic wave propagation in poroelastic media are are described in Ref. [17]. As an example of results, $c_P$ and $c_S$ are shown as a function of $S_H$ in Fig. 1, for a sample having an initial water saturation of $S_{w0}$= 0.73.

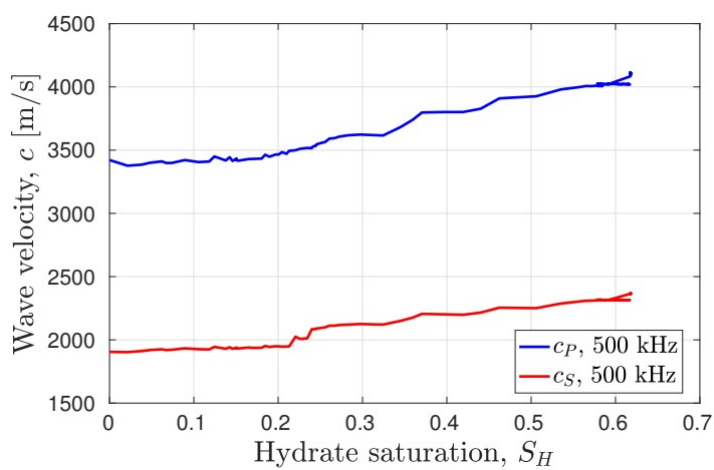

Figure 1: Compressional, $c_P$, and shear $c_S$ wave velocites in a Bentheim sandsone with initial water saturation 0.73 during hydrate growth, $S_H$.

The measurements show that as hydrates form, in the sandstone, the elastic wave velocities increase. A moderate increase is observed for $S_H$<0.2. For $S_H$> 0.2, a more clear increase is seen in $c_P$ and $c_S$. Additional measurements and modeling results are presented and discussed in Ref. [17].

## References

[1] G. J. Moridis, T. S. Collett, R. Boswell, M. Kurihara, M. T. Reagan, C. Koh, and E. Sloan, "Toward production from gas hydrates," Proceedings of SPE Reservoir Evaluation and Engineering, Keystone, Colorado, USA (2009)

[2] G. Moridis, T. S. Collett, M. Pooladi-Darvish, S. H. Hancock, C. Santamarina, R. Boswell, T. J. Kneafsey, J. Rutqvist, M. B. Kowalsky, and M. T. Reagan, "Challenges, uncertainties, and issues facing gas production from gas-hydrate deposits," Proceedings of SPE Reservoir Evaluation and Engineering, Keystone, Colorado, USA (2011)

[3] A. V. Milkov, "Global estimates of hydrate-bound gas in marine sediments: How much is really out there?," Earth-Science Reviews, vol. 66, pp. 183-197 (2003)

[4]   W. F. Waite, W. J. Winters, and D. H. Mason, "Methane hydrate formation in partially water-saturated Ottawa sand," American Mineralogist, vol. 89, pp. 1202-1207 (2004)

[5]   W. J. Winters, I. A. Pecher, W. F. Waite, and D. H. Mason, "Physical properties and rock physics models of sediment containing natural and laboratory-formed methane gas hydrate," American Mineralogist, vol. 89, pp. 1221-1227 (2004)

[6]   J. A. Priest, A. I. Best, and C. R. Clayton, "A laboratory investigation into the seismic velocities of methane gas hydrate-bearing sand," Journal of Geophysical Research: Solid Earth, vol. 110, pp. B04102 (2005)

[7]   S. Nakagawa and T. Kneafsey, "Split Hopkinson Resonant Bar test and its application for seismic property characterization of geological media," Proceedings of the 44th US Rock Mechanics Symposium and 5th US-Canada Rock Mechanics Symposium, American Rock Mechanics Association, Salt Lake City, Utah (2010)

[8]   G. Guerin and D. Goldberg, "Modeling of acoustic wave dissipation in gas hydrate-bearing sediments," Geochemistry Geophysics Geosystems, vol. 6, pp. 2005GC000918 (2005)

[9]   S. Chand, T. A. Minshull, D. Gei, C., and M. José, "Elastic velocity models for gas-hydrate-bearing sediments - A comparison," Geophysical Journal International, vol. 159, pp. 573-590 (2004)

[10]  G. W. Hu, Y. G. Ye, J. Zhang, C. L. Liu, S. B. Diao, and J. S. Wang, "Acoustic properties of gas hydrate–bearing consolidated sediments and experimental testing of elastic velocity models," Journal of Geophysical Research, vol. 115, pp. B02102 (2010)

[11]  M. B. Rydzy and M. L. Batzle, "Ultrasonic velocities in laboratory-formed gas hydrate-bearing sediments," in 23rd EEGS Symposium on the Application of Geophysics to Engineering and Environmental Problems, Society of Exploration Geophysicists and Environment and Engineering Geophysical Society, Keystone, Colorado (2010)

[12]  Q. Zhang, F. G. Li, C. Y. Sun, Q. P. Li, X. Y. Wu, B. Liu, and G. J. Chen, "Compressional wave velocity measurements through sandy sediments containing methane hydrate," American Mineralogist, vol. 96, pp. 1425–1432 (2011)

[13]  J. B. Molyneux and D. R. Schmitt, "First-break timing: Arrival onset times by direct correlation," Geophysics, vol. 64, pp. 1492-1501 (1999)

[14]  J. B. Molyneux and D. R. Schmitt, "Compressional-wave velocities in attenuating media: A laboratory physical model study," Geophysics, vol. 65, pp. 1162-1167 (2000)

[15]  W. Sachse and Y. H. Pao, "On the determination of phase and group velocities of dispersive waves in solids," Journal of Applied Physics, vol. 49, pp. 4320-4327 (1973)

[16]  K. W. Winkler and T. J. Plona, "Technique for measuring ultrasonic velocity and attenuation spectra in rocks under pressure," Journal of Geophysical Research, vol. 87, pp. 10776-10780 (1982)

[17]  M. M. Sæther, "Elastic wave velocities under methane hydrate growth in Bentheim sandstone," PhD Thesis, University of Bergen, Dept. of Physics and Technology, Bergen, Norway (2018)

[18] P. He and J. Zheng, "Acoustic dispersion and attenuation measurement using both transmitted and reflected pulses," Journal of the Acoustic Society of America, vol. 39, pp. 27-32 (2001)

[19] M. Helgerud, J. Dvorkin, A. Nur, A. Sakai, and T. Collett, "Elastic-wave velocity in marine sediments with gas hydrates: Effective medium modeling," Geophysical Research Letters, vol. 26, pp. 2021-2024 (1999)

[20] J. Dvorkin, A. Nur, and H. Yin, "Effective properties of cemented granular materials," Mechanics of Materials , vol. 18, pp. 351–366 (1994)

[21] B. Tohidi, R. Anderson, M. B. Clennell, R. W. Burgass, and A. B. Biderkab, "Visual observation of gas-hydrate formation and dissociation in synthetic porous media by means of glass micromodels," Geology, vol. 29, pp. 867–870 (2001)

[22] W. Waite, B. DeMartin, S. Kirby, J. Pinkston, and C. Ruppel, "Thermal conductivity measurements in porous mixtures of methane hydrate and quartz sand," Geophysical Research Letters, vol. 29, (2002)

[23] A. E. Peksa, K.-H. A. Wolf, and P. L. Zitha, "Bentheimer sandstone revisited for experimental purposes," Marine and Petroleum Geology, vol. 67, pp. 701–719 (2015)